\documentclass{PoS}

\title{Recent progress on nuclear potentials from lattice QCD}

\ShortTitle{Recent progress on nuclear potentials from lattice QCD}

\author{\speaker{Sinya AOKI} \\
        Graduate School of Pure and Applied Sciences, University of Tsukuba, Tsukuba, Ibaraki 305-8571, Japan\\
     Center for Computational Sciences, University of Tsukuba, Tsukuba, Ibaraki 305-8577, Japan\\
        E-mail: \email{saoki@het.ph.tsukuba.ac.jp}}

\author{
for HAL QCD Collaboration\\
\begin{center}
 \includegraphics[ height=0.2\textwidth ]{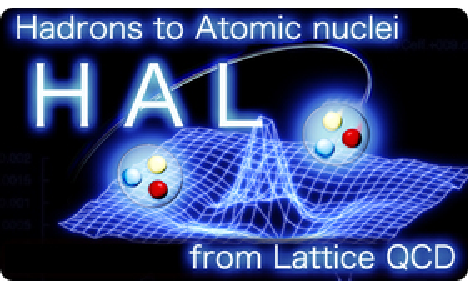} 
\end{center}
}

\abstract{A new method for nuclear potentials is reviewed. The strategy to extract the potential from the Nambu-Bethe-Salpeter wave function in lattice QCD is explained and the first result based on the method is presented in quenched QCD. The recent progress of the method is summarized.}

\FullConference{35th International Conference of High Energy Physics - ICHEP2010,\\
		July 22-28, 2010\\
		Paris France}

\begin{document}

\section{Introduction}
The nuclear force is a basis for understanding the structure of ordinary and hyper nuclei, the structure of neutron stars and the ignition of the Type II SuperNova.
The phenomenological NN potential, described by $O(10)$ parameters to fit $O(1000)$ scattering phase shift data, has the attraction at long to medium distance (the attraction well) as well as the repulsion at short distance (the repulsive core\cite{jastrow}).
Recently a new method has been proposed to extract the potential in lattice QCD and
successfully applied to the NN potential in the quenched simulation\cite{IAH1,AHI1, AHI2}.
In this report we review the strategy and the recent progress of the method.

\section{Strategy in Lattice QCD}
A key quantity for the definition of the potential in QCD is the equal-time Nambu-Bethe-Salpeter(NBS) wave function defined by
\begin{equation}
\varphi_E({\bf r}) = \langle 0 \vert N({\bf x}+{\bf r}, 0) N({\bf x},0)\vert B=2, E\rangle
\end{equation}
where the local nucleon field is constructed from quark fields $q(x)=(u(x), d(x) )$ as 
\begin{equation}
N(x)=\varepsilon_{abc}q^a(x)(u^b(x)C\gamma_5u^c(x))
\end{equation}
and $\vert B=2, E\rangle$  is an eigenstate of QCD with the baryon number 2 and energy $E$.
We consider the center of mass system, so that $E=2\sqrt{{\bf k}^2+ m_N^2}$ with the nucleon mass $m_N$.
If the energy $E$ is below the inelastic threshold $E_{\rm th}=2m_N+m_\pi$ where $m_\pi$ is the pion mass, the NBS wave function becomes\cite{Ishizuka,AHI1}
\begin{equation}
\varphi_E({\bf r}) \simeq \sum_l A_l Y_l(\theta,\phi) \frac{\sin(kr-l\pi/2+\delta_l(k))}{k r},
\quad k=\vert{\bf k}\vert
\end{equation}
in the asymptotic region that $r=\vert{\bf r}\vert \rightarrow \infty$. It is noted that the "phase shift" $\delta_l(k)$ for the partial wave component of the NBS wave function is the phase of the S-matrix in QCD, which is determined by the unitarity of the elastic scattering as $S_l=e^{i2\delta_l}$.

In our strategy\cite{IAH1,AHI1,AHI2} we first define the non-local potential as
\begin{equation}
\left[\epsilon_k - H_0\right]\varphi_E({\bf r}) = \int U({\bf r},{\bf r}^\prime)\varphi_E({\bf r}^\prime)\, d^3r^\prime,
\qquad H_0=\frac{-\nabla^2}{2\mu},\  \epsilon_k =\frac{k^2}{2\mu}
\end{equation}
where $\mu=m_N/2$ is the reduced mass. We then expand $U$ in terms of the velocity(derivative) as $U({\bf r},{\bf r}^\prime)=V({\bf r},\nabla)\delta^3({\bf r} -{\bf r}^\prime)$ where
\begin{equation}
V({\bf r},\nabla) =\underbrace{V_C(r) + V_T(r) S_{12}}_{\rm LO} + \underbrace{V_{\rm LS}(r) {\bf L}\cdot{\bf S}}_{\rm NLO} + O(\nabla^2),\quad S_{12}=\frac{3}{r^2}(\sigma_1\cdot{\bf r})(\sigma_2\cdot{\bf r})-(\sigma_1\cdot\sigma_2).
\end{equation}
Here $\sigma_i/2$ represents the spin of $i$-th nucleon, ${\bf S}=(\sigma_1+\sigma_2)/2$ is the total spin and $S_{12}$ is the tensor operator. In this report we truncated the expansion at the leading order(LO).

\begin{figure}[bt]
\begin{center}
\includegraphics[width=50mm,angle=270,clip]{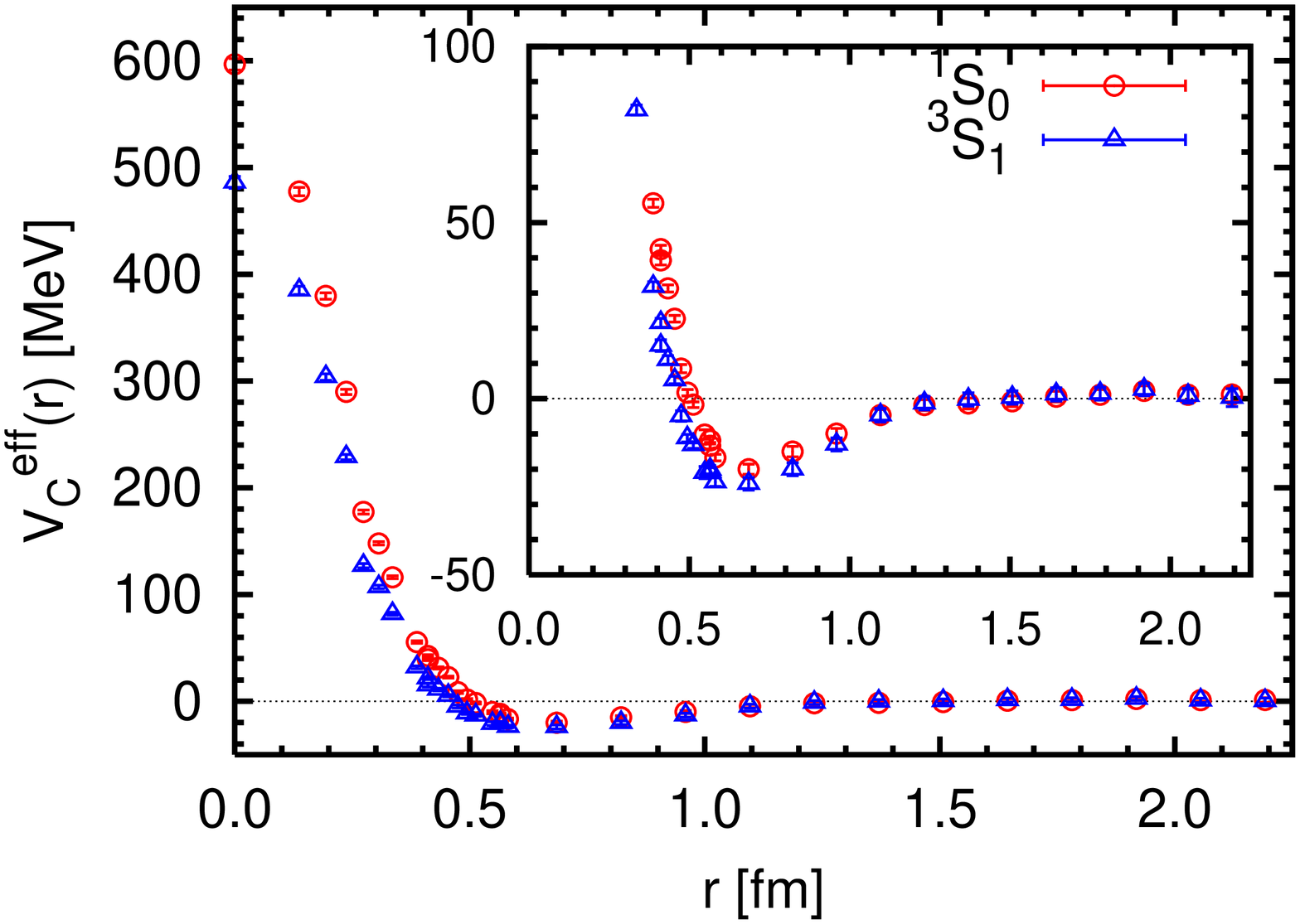} 
\includegraphics[width=50mm,angle=270,clip]{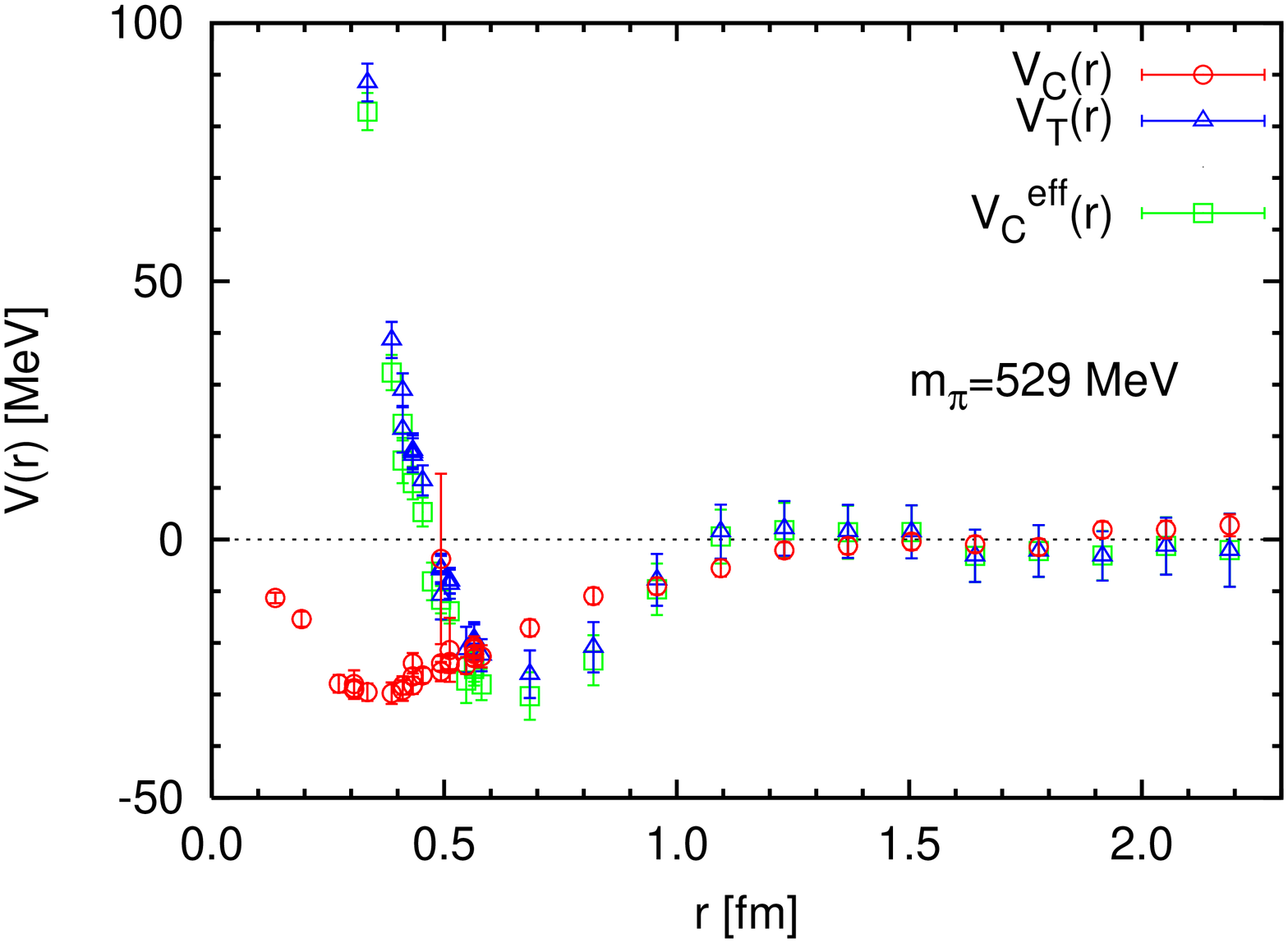} 
\end{center}
\caption{Left: The central (effective central) $NN$ potential (MeV) for the spin singlet (triplet) at $m_\pi\simeq 530$ MeV in quenched QCD\cite{IAH1}.
Right: The tensor potential $V_T(r)$ and the central potential $V_C(r)$, together with the effective central potential, for the spin-triplet at  $m_\pi\simeq 530$ MeV in quenched QCD\cite{AHI1}.}
\label{fig:potential_quench}
\end{figure}
In the left of Fig.~\ref{fig:potential_quench}, we show the first result for the NN central potentials calculated in quenched QCD at $a\simeq 0.14$ fm, $L\simeq 4.4$ fm and  $m_\pi\simeq 530$ MeV\cite{IAH1}.
The potentials reproduce qualitative features of NN potentials, not only the attraction well but also the repulsion core. Ref.\cite{IAH1} has been selected as one of 21 papers in Nature Research Highlight 2007\cite{Nature}.

The validity of  the LO approximation in the derivative expansion for the non-local potential can be examined as follows. If the higher order terms become relevant, the LO local potential depends on the energy $E$ of the NBS wave function. In Ref.\cite{murano1}, the LO potentials obtained at $\epsilon_k \simeq 0$ MeV are compared with theose at $\epsilon_k \simeq 46$ MeV in quenched QCD.  The energy dependence of the LO local potentials turns out to be very small at low energy in our choice of the NBS wave function. This implies that non-locality of $U$ is also weak at low energy.

\section{Recent Developments}
The NBS wave function for the spin-triplet channel has two independent components mixed by the tensor operator $S_{12}$, for example, $^3S_1$ and $^3D_1$. Using these two components, we obtain both central and tensor potentials at the LO. In the right of Fig.~\ref{fig:potential_quench}, the tensor potential as well as the central potential for the spin-triplet channel of $S$ and $D$ waves are shown, in addition to the effective central potential obtained by ignoring the tensor potential. The tensor potential seems not to have the repulsive core, while the difference between the true central potential and the effective central potential is small.

The LO potentials for both spin singlet and triplet channels have been calculated in full QCD at $a\simeq 0.1$ fm and $L\simeq 2.9$ fm\cite{ishii1}.
We observe larger repulsive core for both central potentials as well as larger attraction for the tensor force in full QCD than in quenched QCD. 

\begin{figure}[bt]
\centering
\includegraphics[width=65mm,angle=0, clip]{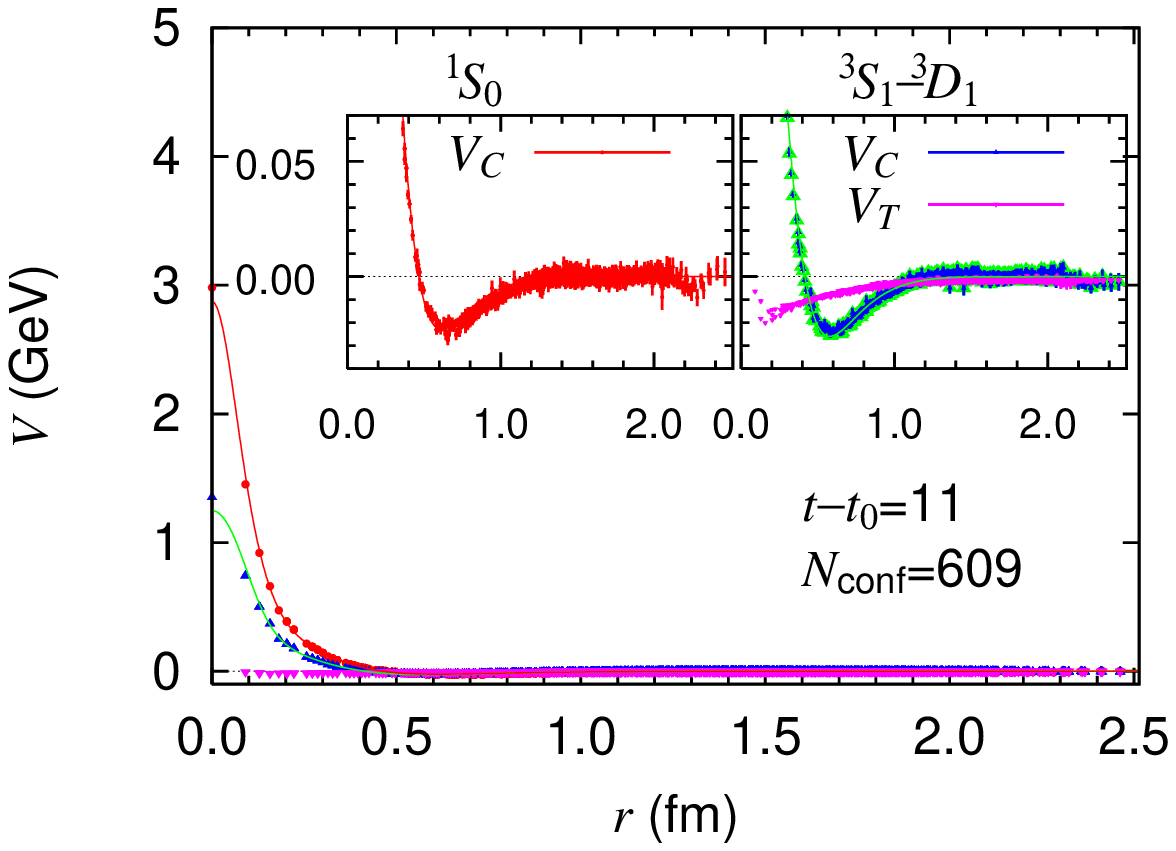} 
\includegraphics[width=60mm,angle=0, clip]{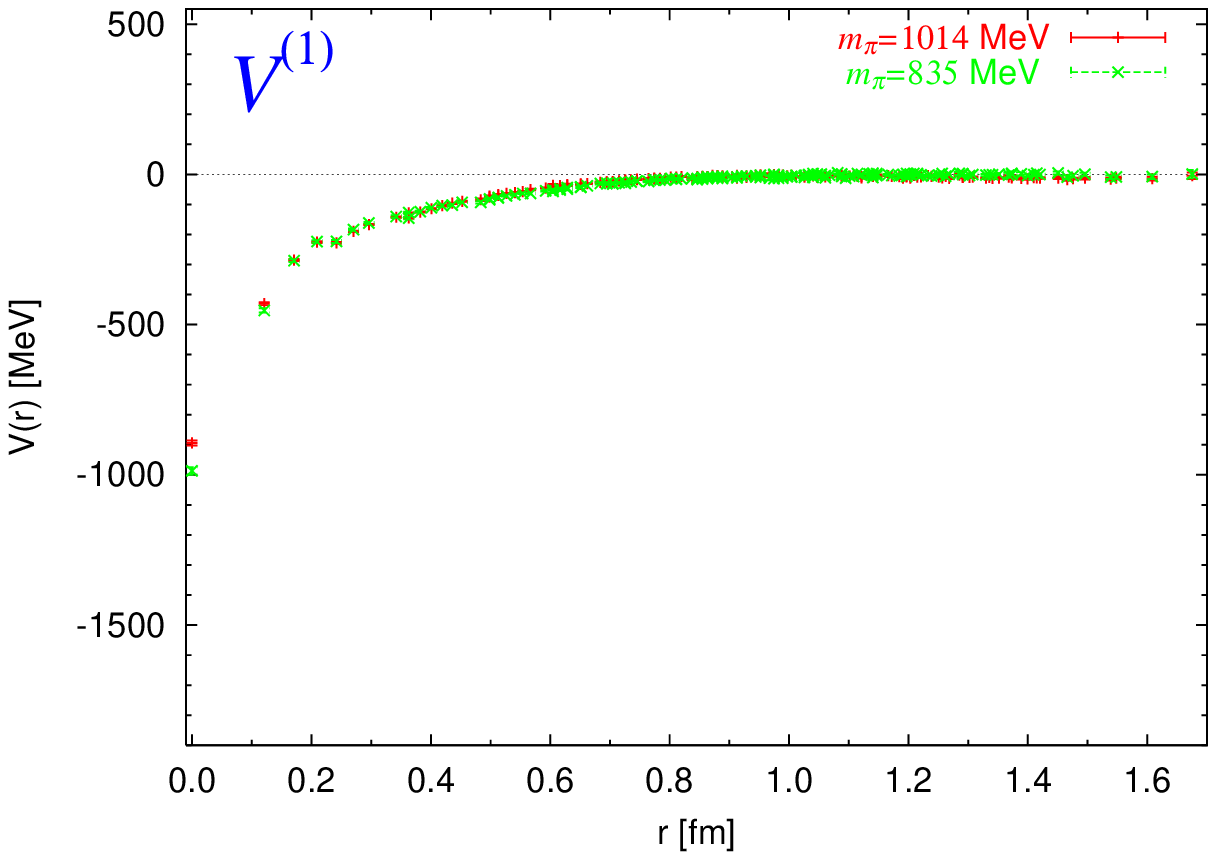} 
\caption{Left: $N\Lambda$ LO potentials in 2+1 flavor full QCD\cite{nemura2}.
Right: The flavor singlet potential in the flavor SU(3) limit at $m_\pi=1014$ MeV (red bars) and 835 MeV (green crosses)\cite{inoue1}.}
\label{fig:hyperon}
\end{figure}
The method can be applied to extractions for hyperon interactions\cite{nemura1,nemura2}. 
The left of Fig.~\ref{fig:hyperon} represents one of such extensions, the $\Lambda N$ potentials in the $I=1/2$ channel, obtained in 2+1 flavor full QCD\cite{nemura2}.
Similar structures found for the NN potential such as the repulsive core and the attractive well have been observed, while the spin dependence of the $\Lambda N$ potentials is larger than that of the NN potentials. We also find that the net interaction is attractive at low energy for both channels.
  
One of the interesting questions is whether the repulsive core is universal or not. To answer this question, interactions between octet baryons have been investigated in the flavor SU(3) symmetric full QCD at $a\simeq 0.12$ and $L\simeq 2$ fm\cite{inoue1}.
The attractive core instead of the repulsive core has appeared for the flavor single potential, as shown in the right of Fig.\ref{fig:hyperon}. This result suggests that the repulsive core is partly related to the Pauli principle among quarks.

\section{Conclusion}
As shown in this report, we can use lattice QCD to extract not only the NN potentials but also more general hadron interactions such as hyperon-nucleon and hyperon-hyperon potentials. 
It will be important to perform such extractions of potentials in full QCD at the physical pion mass,
based on which we can make {\it ab initio} investigates for nuclear and hyper nuclear physics. 

The author  thanks all members of HAL QCD collaboration for useful discussions.
This work is supported in part by the Grant-in-Aid of MEXT( (No.20340047) and by Grant-in-Aid for Scientific Research on Innovative Areas (No.2004: 20105001, 20105003).

\end{document}